# Electrostatic Graphene Loudspeaker

Qin Zhou and A. Zettl[a]

Center of Integrated Nanomechanical Systems, University of California at Berkeley, Berkeley, California 94720, USA

Department of Physics, University of California at Berkeley, Berkeley, California 94720, USA

Materials Sciences Division, Lawrence Berkeley National Laboratory, Berkeley, California 94720, USA

Graphene has extremely low mass density and high mechanical strength, key qualities for efficient wide-frequency-response electrostatic audio speaker design. Low mass ensures good high frequency response, while high strength allows for relatively large free-standing diaphragms necessary for effective low frequency response. Here we report on construction and testing of a miniaturized graphene-based electrostatic audio transducer. The speaker/earphone is straightforward in design and operation and has excellent frequency response across the entire audio frequency range (20HZ – 20kHz), with performance matching or surpassing commercially available audio earphones.

---

[a] Electronic mail: azettl@berkeley.edu

Efficient audio sound transduction has a history dating back millions of years[1]. Primitive insect singers generate loud and pure-tone sound with high efficiency by exciting resonators inside their body[2]. Male crickets generate chirping sounds via stridulation[3], where the scraper edge of one wing is rubbed against the ribbed edge of the other wing. Advantageous structural properties of the wings (relatively large, low-mass flexural membranes) allow extremely efficient muscle-to-sound energy transduction. In a human context, unnatural (i.e. non-voice) sound production has been explored for millennia, with classic examples being drumheads and whistles for long-range communications and entertainment[4]. In modern society, efficient small-scale audio transduction is ever more important for discrete audio earphones and microphones in portable or wireless electronic communication devices.

For human audibility, an ideal speaker or earphone should generate a constant sound pressure level (SPL) from 20 Hz to 20 kHz[5,6], i.e. it should have a flat frequency response. Most speakers available today reproduce sound via a mechanical diaphragm, which is displaced oscillatorily during operation. The diaphragm, with inherent mass, restoring force (i.e. spring constant), and damping, essentially constitutes a simple harmonic oscillator. Unlike most insect or musical instrument resonators which exhibit lightly-damped sharp frequency response, a wide-band audio speaker typically requires significant damping to broaden the response. Unfortunately, "damping engineering" quickly becomes complex and expensive, with inevitable power inefficiencies.

An alternative approach to response spectrum broadening is to reduce both the mass and spring constant of the diaphragm so that inherent air damping dominates and flattens the response peaks. Moreover, with ambient air serving as the dominant damping mechanism, most input energy is converted to a sound wave, which makes such speakers highly power efficient. For these reasons (more detailed analysis is provided in the Supporting Information), the ideal audio transduction diaphragm should have small mass and a soft spring constant, and be non-perforated to efficiently displace the surrounding air. Electrostatically-driven thin- membrane loudspeakers[7] employing an electrically conducting, low-mass diaphragm with significant air damping have been under development since the 1920's (the first were made of pig intestine covered with gold-leaf), but miniaturized electrostatic earphones are still rare. One key reason is that the per-area air damping coefficient significantly decreases when the size of the diaphragm falls below the sound wavelength[8]. Hence, for small speakers a thinner and lower mass density diaphragm is required to continue the dominance of air damping. Such a diaphragm is difficult to realize. If conventional materials such as metalized mylar are made too thin, they invariably fatigue and break.

Graphene is an ideal building material for small, efficient, high-quality broad-band audio speakers because it satisfies all the above criteria. It is electrically conducting, has extremely small mass density[9], and can be configured to have very small effective spring constant. The effective spring constant of a thin circular membrane is[10]

$$k_{eff} = 4\pi\sigma t \quad (1)$$

where $\sigma$ is the stress and $t$ is the thickness of the membrane. It is convenient to use per-area values for modeling the diaphragm vibration (see Su Information) since for a given membrane the mass per unit area is constant. The equivalent per area spring constant is therefore

$$k = \frac{k_{eff}}{Area} = \frac{4\sigma t}{R^2} \quad (2)$$

where $R$ is the radius of the circular membrane. We note that the spring constant $k$ scales proportionally with the thickness of the membrane and inversely with the 2$^{nd}$ power of the radius of the membrane. The exceptional mechanical strength of graphene[11] makes it possible to construct large and thin suspended diaphragms, which effectively reduces *k*.

Graphene was previously used to construct a thermoacoustic loudspeaker[12-14]. In the thermoacoustic configuration graphene serves as a stationary heater to alternately heat the surrounding air thereby producing, via thermal expansion, a time-dependent pressure variation, i.e. sound wave. The method is especially effective in the ultrasonic region because of graphene's small heat capacity (for this reason, carbon nanotube films can also be utilized[15-17]). However, for thermoacoustic speakers operating at audio frequencies, most input energy is dissipated by heat conduction through the air and does not generate sound[12]. For example, the power efficiency for a graphene thermoacoustic speaker is exceedingly small, decreasing from ~$10^{-6}$ at 20 kHz to ~$10^{-8}$ at 3 kHz[12,13]. The thermoacoustic approach also suffers from sound distortion because the heating power is proportional to the square of the input signal and the transduction is therefore intrinsically non-linear[15].

We here describe an electrostatically driven, high-efficiency, mechanically vibrating graphene-diaphragm based audio speaker. Even without optimization, the speaker is able to produce excellent frequency response across the whole audible region (20 Hz~20 KHz), comparable or superior to performance of conventional-design commercial counterparts.

Figure 1 shows schematically the electrostatically driven graphene speaker (EDGS). A multilayer graphene membrane is suspended midway between two actuating perforated electrodes. The graphene is DC biased at $V_{DC}$. With no input signal, the electrostatic forces from the upper and lower electrodes balance. When the two driving electrodes are driven with opposite polarity at $V_{in}$, the total electrostatic force applied on graphene is (per unit area)

$$F = F_1 - F_2 = \frac{\varepsilon}{2d^2}(V_{DC} + V_{in})^2 - \frac{\varepsilon}{2d^2}(V_{DC} - V_{in})^2 = \frac{\varepsilon V_{DC}}{d^2}V_{in} \qquad (3)$$

where $F_1$ and $F_2$ are force magnitudes due to the respective electrodes, $\varepsilon$ is the electric permittivity of air, and $d$ is the nominal distance between graphene and electrodes. Eq. (3) shows that the actuating force is linearly proportional to the input signal, a key advantage for low sound distortion. A protective $SiO_2$ layer is deposited on the electrodes to prevent the graphene from accidentally shorting to the electrodes at very large drive amplitude.

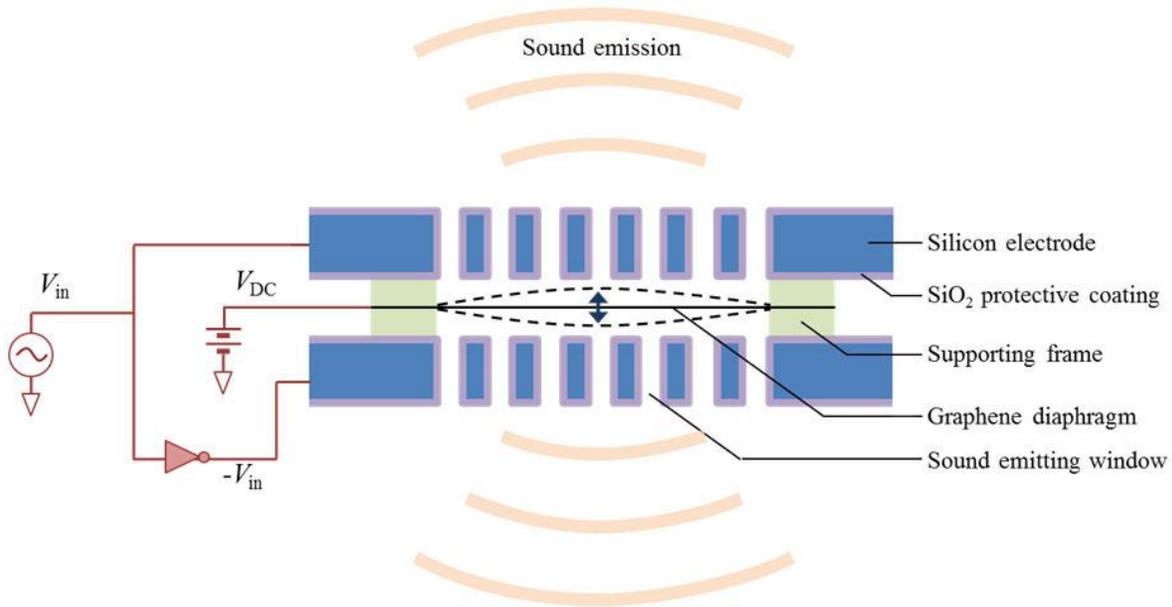

Figure 1. Schematics of the graphene-based EDGS speaker. A graphene diaphragm, biased by a DC source, is suspended midway between two perforated electrodes driven at opposite polarity. The varying electrostatic force drives the graphene diaphragm which in turn disturbs air and emits sound through the electrodes. The light mass and low spring constant of the graphene diaphragm, together with strong air damping, allow for high-fidelity broad-band frequency response. Such a speaker also has extremely high power efficiency.

To fabricate the EDGS structure the graphene is first attached to a suspension frame (Figure 2a), which is then sandwiched between separately fabricated electrodes (Figure 2b). Multilayer graphene is synthesized on 25-μm-thick nickel foil in a CVD furnace at 1000 °C[18-20]. The foil is first annealed at 1000 °C for 1 hour with 50 sccm hydrogen flow at 200 mTorr, after which the hydrogen flow is increased to 100 sccm and methane is introduced at 5 sccm to start the growth process. The growth pressure is 2 Torr. After 20 minutes, the furnace is turned off and the nickel foil is quickly removed from the hot zone to allow the formation of graphene layers. After the growth, a self-adhesive circular frame (Avery® ETD-909, 60 μm thick) with 7 mm diameter opening is attached on the nickel foil. The foil is then etched away with 0.1 g/ml $FeCl_3$ solution, so that the graphene membrane is only attached to and supported by the circular frame. The frame is first transferred to fresh DI water bath several times to clean the etchant residue, and then immersed in acetone. We find that the multilayer graphene diaphragm is strong enough to be directly dried in air by pulling the frame out from acetone.

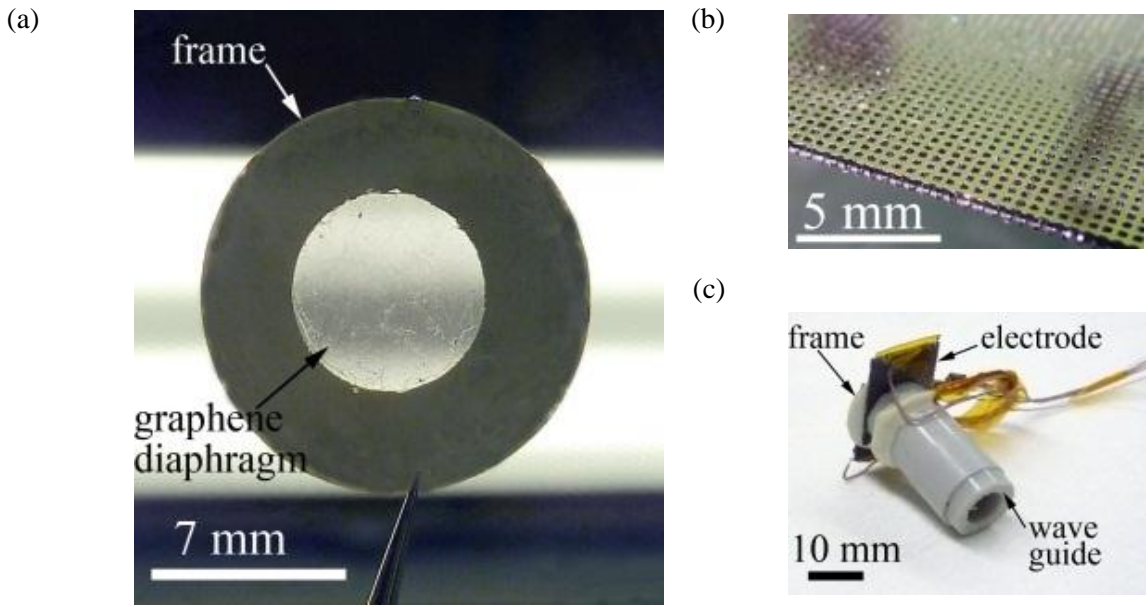

Figure 2. Images of (a) 7mm diameter graphene diaphragm suspended across annular support frame, (b) actuating electrodes, and (c) assembled speaker.

The thickness of the free-standing graphene diaphragm is determined by light transmission measurement to be ~30 nm (22%~25% transmission). The electrical contact to the graphene diaphragm is made by attaching a 20-μm-diameter gold wire to the portion of graphene lying on the supporting frame. Another circular frame is attached to the original frame from the opposite side (so that the graphene diaphragm is sandwiched between them) to fix the gold wire. The frames also serve as spacers between graphene and the electrodes in the speaker assembly. The gap distance can be increased by stacking multiple (empty) frames on each other. For the results presented here we use two frames on each side of the graphene, which gives a gap distance $d$ = 120 μm.

The electrodes are constructed from silicon (525 μm thick, resistivity 10 Ohm·cm, test grade). A photolithography and deep-reactive-ion-etching step are used to construct through-wafer square holes of 250 μm wide as sound emitting windows. A 500 nm protective wet thermal oxide layer is then grown on the wafer at 1050 ºC (Figure 2b). The wafer is then diced as electrodes. Dicing the wafer also exposes the silicon so that electric connections are made by attaching conductive wires to the edges of the electrodes with silver paste.

For prototype demonstration, two electrodes and one graphene diaphragm are simply sandwiched together and held by a spring clip. In another implementation, a 7 mm inner-diameter pipe, serving as a wave guide, is perpendicularly attached to the surface of the electrodes to facilitate sound coupling between the speaker and a listener's ear (Figure 2c). This improves far-field efficiency for a small speaker operating at wavelength larger than the diaphragm size.

We now describe performance tests for the EDGS speaker. $V_{DC}$=100 V is used to bias the device, and the input sound signal is introduced from a signal generator or from a commercial laptop or digital music player. The maximum amplitude of the input signal $V_{in}$ used in the test is 10V. The operation current is usually a few nano-amps, indicating very low power consumption (<<1 μW) and high power efficiency. In fact, the power efficiency of an electrostatic speaker can be exceedingly high (close to 1) because the power dissipation path is almost pure air damping[7], which converts the mechanical vibration of diaphragm to sound. Magnetic coil type earphones (the type used today for virtually all earphone applications) typically have efficiencies <0.1.

The sound generated by the graphene speaker is easily audible by the human ear. The fidelity is qualitatively excellent when listening to music. To quantitatively characterize the speaker, the frequency response curve is measured from 20 Hz to 20 kHz and compared to a commercial earphone of similar size (Sennheiser® MX-400). The sound card in a laptop computer (SoundMAX® Integrated Digital HD Audio) is used to generate equal-amplitude sine waves, and a commercial condenser microphone (SONY® ICD-SX700) is used to measure the sound pressure level (SPL) at different frequencies. The software is Room EQ Wizard.

Figure 3 shows the sound pressure level over the relevant audio frequency range for the EDGS speaker (Figure 3a), the Sennheiser® MX-400 (Figure 3b), and a miniature thermoacoustic speaker (Figure 3c, adapted from Ref. 13). The graphene speaker, with almost no specialized acoustic design, performs comparably to a high quality commercial headset. Moreover, the high-frequency performance of the EDGS (Figure 3a) is markedly better than that for the MX-400 thanks to the extremely low-mass diaphragm. In the low frequency region, the EDGS and MX-400 response curves both bend downward, largely due to limited capability of the sensing microphone and restricted coupling between the speaker and microphone. Even so, the low-frequency performance of the EDGS speaker is markedly superior to that predicted for a thermoacoustic speaker (dashed line, Figure 3c)). At very high frequencies (>10kHz), the thermoacoustic speaker maintains its excellent high frequency response, but, as mentioned above, the power efficiency is at least six orders of magnitude lower than that for the EDGS, which makes it impractical for most portable applications.

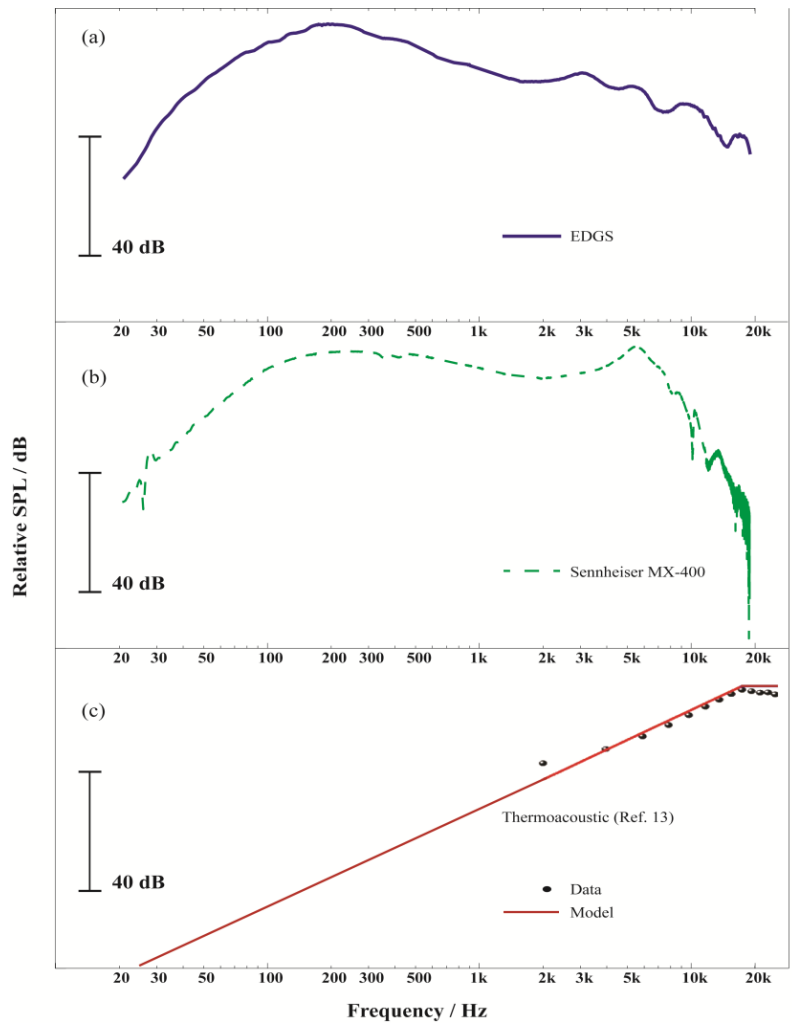

Figure 3. Frequency response of various miniature audio speakers. a) Graphene diaphragm EDGS speaker (this work); b) Commercially available Sennheiser® MX-400 magnetic coil speaker; c) Thermoacoustic speaker. The points in panel c) are experimental data from ref. 13,

while the solid red line is the theoretically predicted behavior for an ideal thermoacoustic speaker. The EDGS speaker performs noticeably better than the commercial voice-coil speaker at high frequencies, both in terms of maintaining high response and avoiding sharp resonances (the slow oscillations in the EDGES curve are due to sound wave interference in the space between the speaker and microphone and they depend on the relative position of the speaker and microphone, but the main trend is consistent). In the low frequency region, both EDGES and the MX-400 perform well, while the thermoacoustic response falls precipitously already below 15kHz. The decrease in the response curves in a) and b) at very low frequency are largely due to limited capability of the microphone and the inefficient coupling between the speaker and microphone.

The speaker-to-microphone performance test has limited accuracy, because the measured response curve is for the whole system - from the sound card to amplifier, to speaker, to microphone, and finally back to sound card. Every transduction introduces some distortion in the measurement. For example, the response is sensitive to the relative position between the speaker and the microphone. Since the focus of this letter is not the detailed acoustic design of a complete sound system but rather the capability of the EDGS graphene diaphragm, we employ laser Doppler velocimetry (LDV) to directly measure the mechanical response limits of the diaphragm. The measured frequency response is illustrated in Figure 4. Within experimental error the LDV frequency response curve for the EEGS diaphragm is relatively flat from 20Hz to 20kHz, which is the desired response of an ideal speaker diaphragm.

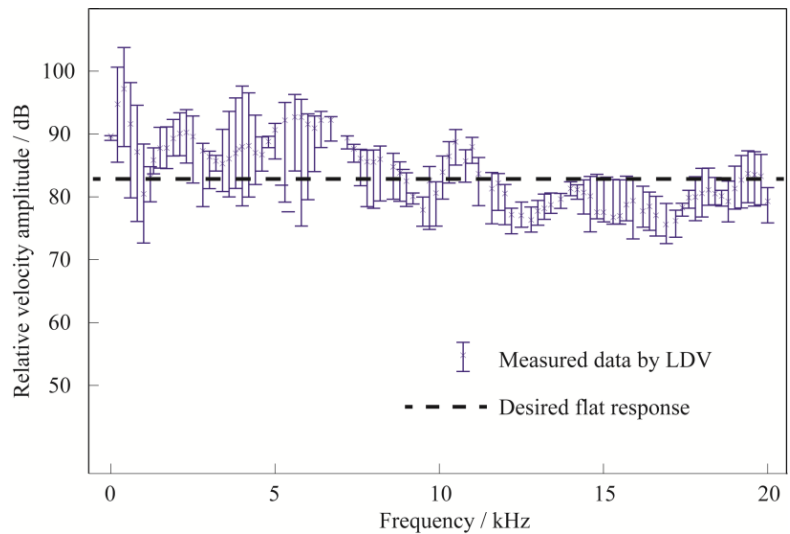

Figure 4: Vibration velocity of graphene diaphragm in EDGS v.s. frequency, measured by Laser Doppler Velocimetry (LDV). Such a measurment is useful because it eliminates extrinsic effects (e.g. acoustic structural design, sound card, microphone response), and represents the "pure" response of the graphene diaphragm itself. Within the error limit of the LDV setup, the response curve appears to be quite flat, indicating that graphene serves as an ideal key component for loudspeakers.

In summary we have demonstrated a robust speaker built from multi-layered graphene diaphragm. The diaphragm is driven electrostatically and reproduces sound with high fidelity. The CVD grown graphene on nickel foil is straightforward and the technique can be easily scaled to construct larger speakers by arraying the graphene diaphragm. The configuration described in this letter could also serve as a microphone. The microphone should also have outstanding response characteristics due to the graphene's ultra-low mass and the excellent coupling to ambient air.

This work was supported in part by the Director, Office of Energy Research, Office of Basic Energy Sciences, Materials Sciences and Engineering Division, of the U.S. Department of Energy under Contract No. DE-AC02-05CH11231, which provided for graphene growth and characterization; by the Office of Naval Research under grant No. N00014-09-1066, which provided for graphene transfer and electrode manufacture, and by the National Science Foundation under Grant No. EEC-083819, which provided for design, construction, and testing of the device. The authors thank Yung-Kan Chen and Prof. David Bogy for assistance with LDV measurements.

APPENDIX: Modeling a loudspeaker using harmonic oscillator.

To examine why the ideal diaphragm for speakers should have small mass and spring constant, we model the diaphragm as a 2$^{nd}$ order spring-damping-mass system, where the mass comes from the diaphragm (as well as the actuating element if it is vibrating together with the diaphragm, which is often true for most speakers where the force-generating coil is directly attached to the diaphragm), the damping force comes from the air, and the spring represents the restoring force that brings the diaphragm to the balanced position.

The equation that describes the movement of the diaphragm is given by

$$m\ddot{x} + \zeta \dot{x} + kx = F \qquad (S1)$$

where $m$ is the mass, $\zeta$ is the damping coefficient, $k$ is the spring constant, and $F$ is the driving force applied on the diaphragm. When driving by a sinuous signal at frequency $\omega$, the vibration amplitude is

$$|\dot{x}| = \frac{|F|}{|\zeta - i\omega^{-1}k + i\omega m|} \qquad (S2)$$

Here the vibration amplitude is represented in terms of velocity rather than displacement, because SPL is directly determined by the velocity amplitude of air:

$$SPL = c\rho |\dot{x}| \qquad (S3)$$

where $c$ is the sound velocity and $\rho$ is the mass density of air. From Eq. (S2), we find that large $k$ results in poor low frequency response (the $\omega^{-1}$ term in denominator), while large m results in poor high frequency response (the $\omega$ term in denominator). As a result, to maintain constant SPL across the whole frequency region, the spring constant $k$ and the mass of the diaphragm $m$ should both be significantly smaller than the air damping coefficient $\zeta$ (strictly speaking, the criteria should be $\omega_{max}^{-1}k \ll \zeta, ?\omega_{min}m \ll \zeta$, where $\omega_{max}$ and $\omega_{min}$ represent the upper and lower bound of the interested frequency region).